\documentclass[aps,twocolumn,showpacs,amsmath,amssymb]{revtex4}
\usepackage{amsthm}
\usepackage{latexsym}
\usepackage{amsfonts}
\usepackage{bbm,dsfont}

\newtheorem{theorem}{Theorem}

\newcommand{\hi}{\mathcal{H}} 

\newcommand{\tr}[1]{\mathrm{tr}\left[#1\right]} 
\newcommand{\ket}[1]{|#1\rangle} 
\newcommand{\kb}[2]{|#1\,\rangle\langle\,#2|} 

\newcommand{\Po}{\mathsf{P}} 
\newcommand{\Ecan}{\Phi} 

\def\<{\langle}
\def\>{\rangle}
\def\d{{\mathrm d}}

\begin{document}

\title{The Canonical Phase Measurement is Pure}

\author{Teiko Heinosaari}
\email{heinosaari@nbi.dk}
\affiliation{Niels Bohr Institute, Blegdamsvej 17, 2100 Copenhagen, Denmark}

\author{Juha-Pekka Pellonp\"a\"a}
\email{juhpello@utu.fi}
\affiliation{Department of Physics and Astronomy, University of Turku, FI-20014 Turku, Finland}

\begin{abstract}
We show that the canonical phase measurement is pure in the sense that the corresponding positive operator valued measure (POVM) is extremal in the convex set of all POVMs.  This means that the canonical phase measurement cannot be interpreted as a noisy measurement, even if it is not a projection valued measure.
\end{abstract}


\maketitle

All the basic building blocks of quantum mechanics, namely states, channels and measurements, have convex structures. A convex combination corresponds to a classical randomization, or mixing, between two or more alternatives. An extremal element does not, by definition, have any (non-trivial) convex decompositions. Extremal elements are thus free from classical noise and for this reason they are called \emph{pure}.

The convex structure of quantum states is important but also very straightforward. If the mathematical description of a state $\varrho$ (i.e. the density operator) is given, then one can calculate the purity $\tr{\varrho^2}$ and the pure states correspond to the maximal value $\tr{\varrho^2}=1$. Hence, the pure states are mathematically described by unit vectors. Also the convex structure of quantum channels is well understood and a simple characterization of pure quantum channels has been derived \cite{Choi}. 

Compared to these two cases of states and channels, the convex structure of measurements is somewhat more complicated. Although the mathematical description of quantum measurements as positive operator valued measures (POVMs) is efficient and elegant, it is not so concise as the corresponding mathematical descriptions of states and channels. 

For a finite dimensional system, a criterion for extremality was given already in \cite{Stormer} if a measurement has a finite number of measurement outcomes. In a recent work of Chiribella et al.\ \cite{ChDASc} it was shown that in finite dimension all pure measurements are concentrated on a  finite number of outcomes. Therefore, a full characterization of pure measurements in finite dimension is obtained. 

If the dimension of the Hilbert space is infinite, the extremality question for quantum measurements is more complicated and there are few results on the characterization of extremal measurements. As the infinite dimensional Hilbert space is used in quantum optics and in continuous variable quantum information, this question is of great importance. Let us notice that the above mentioned extremality criteria for states and channels do not depend on the dimension of the Hilbert space.
It is therefore interesting that finite vs. infinite dimension makes a fundamental difference in the structure of quantum measurements. Namely, in contrast to finite dimension in infinite dimension there are extremal measurements which are neither sharp nor discrete.

Let us briefly recall the mathematical description of quantum measurements via \emph{positive operator valued measures} (POVMs) \cite{OQP97}. Consider a quantum system with Hilbert space $\hi$ and suppose that the measurement outcomes form a set $\Omega$. The set $\Omega$ can be finite or infinite, depending on the given measurement. A POVM is a function $\Po$ which associates to each subset $X\subseteq\Omega$ a positive operator $\Po(X)$ acting on $\hi$ \cite{sigma}. It is required that for every state $\varrho$, the mapping
\begin{equation*}
X\mapsto p_\varrho(X) = \tr{\varrho\Po(X)}
\end{equation*}
is a probability distribution. Especially, $\Po$ satisfies a normalization condition $\Po(\Omega)=I$. 
The number $p_\varrho(X)$ is the probability of getting a measurement outcome $x$ belonging to $X$, when the system is in the state $\varrho$ and the measurement $\Po$ is performed. 

A POVM $\Po$ is a \emph{projection valued measure} (PVM) if $\Po(X)^2=\Po(X)$ for all $X\subseteq\Omega$.  Projection valued measures are usually identified with \emph{sharp} quantum measurements. It is easy to see that all sharp measurements are pure \cite{PSAQT82}. In a finite dimensional Hilbert space all sharp measurements are supported only on a finite set. However, if $\hi$ is infinite dimensional we clearly have also sharp measurements with continuous spectra, e.g. position and momentum.
 
There are also other pure measurements than the sharp ones \cite{PSAQT82} and various examples in finite dimension are given in \cite{DaLoPe05}. Let us notice that a POVM defined on a finite dimensional Hilbert space can be extend to infinite dimension by adding one more outcome and normalizing the POVM accordingly. In this process pure measurements clearly stay pure.  More generally, we can split the infinite dimensional Hilbert space into finite dimensional subspaces and define pure measurements in each subspace separately. The total collection then forms a pure measurement in the infinite dimensional space. In this way we can get pure measurements which are \emph{discrete}, i.e., have an either finite or countably infinite number of outcomes. 

We will now show that in infinite dimension there is a pure quantum measurement which is neither sharp nor discrete, and which has no direct link or connection to these two classes, having e.g.\ no interpretation as an approximate joint measurement of sharp quantities  \cite{holevo}. 
Moreover, our example, the \emph{canonical phase measurement}, is not an artificial mathematical construction but it is an important object in quantum optics (see e.g.\ \cite{dis} and references therein). 

It is generally accepted that the canonical phase measurement for the single-mode radiation field is represented by the \emph{London phase distribution}. 
Hence, the canonical phase measurement has the POVM $\Ecan$ defined by
\begin{equation}\label{eq:ecan}
\Ecan(X)=\sum_{n,m=0}^\infty\frac{1}{2\pi}\int_Xe^{i(n-m)\theta}\mathrm{d}\theta \ \kb{n}{m} \, .
\end{equation}
Here $X\subseteq [0,2\pi)$ and $\ket{n}$ is the number basis. The canonical phase measurement can also be represented using the \emph{Susskind-Glogower phase states} $\ket{\theta}=\sum_{n=0}^{\infty} e^{in\theta} \ket{n}$ by
\begin{equation}\label{eq:ecan-sg}
\Ecan(X)= \frac{1}{2\pi}\int_X \ \kb{\theta}{\theta} \ \mathrm{d}\theta \, .
\end{equation}
Note that the phase states $\ket{\theta}$ are generalized vectors so that they do not belong to the Hilbert space (spanned by the number states) and the above equation must be understood as a sesquilinear form \cite{gen}.

We further recall that the canonical phase measurement arises as the limiting distribution of the \emph{Pegg-Barnett formalism}.
In addition, $\Ecan$ has been independently derived by Holevo \cite{PSAQT82} and Helstrom \cite{QDET76} in the more general framework of quantum estimation theory.
We would like to emphasize that the canonical phase measurement is not a slight deviation from a sharp measurement but it has some very different qualitative properties \cite{properties}. 

To understand why POVMs inevitably arise in the description of phase measurements, recall that in quantum optics coherent states describe laser light and any coherent state has a natural phase parameter (the phase parameter of a coherent state $\ket{z}$ is arg$(z)$). 
Phase probability distributions related to two coherent states with same amplitude but different phase parameters should differ only by the difference of the phase parameters; otherwise the measurement can hardly be consider a \emph{phase} measurement.
This leads to a requirement that a phase measurement should be described as a POVM which is \emph{phase shift covariant} in the sense that a phase shifter 
shifts the phase distribution without changing its shape \cite{shift}.

There exist an infinite number of phase shift covariant POVMs, the most important being the canonical phase measurement $\Ecan$.
However, it is well known that there is no phase shift covariant PVM.
One could, however, still try to dispute these facts by claiming that  the canonical phase measurement $\Ecan$ is a randomization over some collection of PVMs. 
Our main result, implying that this is not the case, is the following \cite{covariant}.

\begin{theorem}\label{lause}
The canonical phase measurement is pure. 
\end{theorem}

Before going to the proof of Theorem \ref{lause}, let us recall some simple but useful facts on convex decompositions. 
First of all, if a POVM $\Po$ has a convex decomposition $\Po=t \Po_1 + (1-t) \Po_2$ with $0<t<1$, then it also has a convex decomposition of the form $\Po=\frac12 \Po'_1 + \frac12 \Po'_2$. Namely, we can choose a number $0 < \epsilon < \min (t, 1-t)$ and set $\Po'_1 = (t + \epsilon) \Po_1 + (1-t - \epsilon) \Po_2$ and $\Po'_2 = (t - \epsilon) \Po_1 + (1-t + \epsilon) \Po_2$. 

Moreover, we can have more general convex decompositions $\Po = \sum_i t_i \Po_i$, where the sum can be even infinite. This, however, gives rise to a two element convex decomposition since
\begin{equation}
\Po = \sum_{i=1}^N t_i \Po_i = t_1 \Po_1 + (1-t_1) \left( \sum_{i=2}^{N} \frac{t_i}{1-t_1} \Po_i \right) \, .
\end{equation} 

Finally, we can form even continuous convex decompositions 
\begin{equation}
\Po = \int_{-\infty}^{\infty} \Po_x t(x) \d x \, , 
\end{equation} 
where $t(x)$ is a probability density and each $\Po_x$ is a POVM. But, again, we can split this into a sum of two terms
\begin{equation*}
\Po = \bar{t} \int_{-\infty}^{a} \Po_x t(x) \bar{t}^ {-1} \d x +  (1-\bar{t}) \int_{a}^{\infty} \Po_x t(x)  (1-\bar{t})^{-1} \d x \, , 
\end{equation*} 
where $\bar{t}=\int_{-\infty}^{a} t(x) \d x$ and $a$ is chosen in a way that $0<\bar{t}<1$.

In conclusion, if a POVM $\Po$ is not pure, then we can write it as a convex decomposition with two different POVMs and coefficients $\frac12$. 
Even if more general convex decompositions are possible, they give always rise also to this kind of simple decomposition \cite{decomposition}.

\begin{proof}
Assume that $\Ecan=\frac12 \Po_1 + \frac12 \Po_2$ for some POVMs
$\Po_k$, $k=1,2$, with the outcome space $[0,2\pi)$. We need to show that this convex decomposition is trivial, i.e., $\Po_1=\Ecan$.

Since $\frac12\Po_k(X) \le  \Ecan(X)$ for all $X\subseteq[0,2\pi)$ and $\Ecan$ has a (matrix valued) density with respect to $\d\theta$, also $\Po_k$ has a (matrix valued) density with respect to $\d\theta$. 
Indeed, following Lemma 4.1. in \cite{HyPeYl} one can show that any POVM which is absolutely continuous with respect to a positive measure, has a positive semidefinite matrix valued density. 
Especially, $\Po_k$ has the form 
\begin{equation}
\Po_k(X)=\sum_{n,m=0}^\infty\frac1{2\pi}\int_X g^{(k)}_{nm}(\theta)e^{i(n-m)\theta}\d\theta\ \kb{n}{m} \, ,
\end{equation}
where $\big(g_{nm}^{(k)}(\theta)\big)_{n,m=0}^\infty$ is a positive semidefinite complex matrix for every $\theta\in[0,2\pi)$. The normalization $\Po_k\big([0,2\pi)\big)=I$ implies that
\begin{equation}
\frac1{2\pi}\int_0^{2\pi} g^{(k)}_{nm}(\theta)e^{i(n-m)\theta}\d\theta=\delta_{n,m} \, .
\end{equation}
Moreover, since $\Ecan=\frac12 \Po_1 + \frac12 \Po_2$, one gets
\begin{equation}\label{ehto}
\frac12 g_{nm}^{(1)}(\theta)+\frac12 g_{nm}^{(2)}(\theta)=1 
\end{equation}
for all $\theta\in[0,2\pi)\setminus O=:O^{\rm c}$ where $O$ is a set of measure zero.

Let $s$ be a positive integer and ${\bf 1}_s$ an $s\times s$--matrix whose all matrix elements are equal to one.
Since ${\bf 1}_s$ is symmetric it can be diagonalized: $U_s^*{\bf 1}_s U_s=D_s$ where
$U_s$ is a unitary (indeed, an orthogonal real) matrix and $D_s={\rm diag}(\alpha_1,...,\alpha_s)$ is a diagonal matrix. 
But since ${\bf 1}_s$ is a rank-one matrix, it has only one nonzero eigenvalue, say, $\alpha_1$.
Immediately one sees that $\alpha_1=s$ and the associated eigenvector is, for example, $(1,1,...,1)$. 

Let $\theta\in O^{\rm c}$ and define an $s\times s$--matrix $G^{(k)}_s(\theta):=\big(g_{nm}^{(k)}(\theta)\big)_{n,m=0}^{s-1}$. From equation \eqref{ehto} one sees that $\frac12G^{(1)}_s(\theta)+\frac12G^{(2)}_s(\theta)={\bf 1}_s$ and, hence,
\begin{equation}\label{tokaehto}
\frac12\tilde G^{(1)}_s(\theta)+\frac12\tilde G^{(2)}_s(\theta)={\rm diag}(s,0,0,...,0) \, ,
\end{equation}
where the matrices $\tilde G^{(k)}_s(\theta)\equiv \big(\tilde g_{nm}^{(k)}(\theta)\big)_{n,m=0}^{s-1}:=U_s^*G^{(k)}_s(\theta)U_s$ are positive semidefinite. Especially,
$\tilde g^{(k)}_{nn}(\theta)\ge 0$, so that equation \eqref{tokaehto} implies that
$\tilde g^{(1)}_{nn}(\theta)=0=\tilde g^{(2)}_{nn}(\theta)$ for all $n=1,...,s-1$.
Moreover, since $|\tilde g^{(k)}_{nm}(\theta)|^2\le \tilde g^{(k)}_{nn}(\theta)\tilde g^{(k)}_{mm}(\theta)$ by positivity, it follows that $\tilde g^{(k)}_{nm}(\theta)=0$ when $(n,m)\ne(0,0)$. Thus,
$
\tilde G^{(1)}_s(\theta)=\lambda_s(\theta){\rm diag}(s,0,0,...,0)
$
where $\lambda_s$ is some (unknown) non-negative function. Hence,
\begin{equation}
G^{(1)}_s(\theta)=U_s\tilde G^{(1)}_s(\theta)U_s^*=\lambda_s(\theta){\bf 1}_s \, .
\end{equation}
Obviously, the functions $\lambda_s$ cannot depend on $s$. Therefore, $\lambda_s=\lambda$ for all $s=1,2,...$, where $\lambda$ is a non-negative function on $[0,2\pi)$. We thus get
\begin{equation}
\Po_1(X)=\int_X\lambda(\theta)\ \d\Ecan(\theta) \, ,
\end{equation}
and the normalization $\Po_1([0,2\pi))=I$ equals to
\begin{equation}
\frac{1}{2\pi}\int_0^{2\pi}\lambda(\theta)e^{i(n-m)\theta}\d\theta=\delta_{n,m} \, .
\end{equation}
But $X\mapsto\frac{1}{2\pi}\int_X\lambda(\theta)\d\theta$ is a probability measure on the unit circle and probability measures are fully determined by their Fourier coefficients. Therefore, $\lambda(\theta)=1$ for almost all $\theta\in[0,2\pi)$ which means that $\Po_1=\Ecan$.
\end{proof}

The fact that the canonical phase measurement is pure demonstrates its fundamental character. 
In particular, it cannot be observed by measuring several observables and then mixing their statistics. 
It follows from our earlier discussion that it can neither be reduced to a continuous random mixture of some other observables.
There are also other known facts supporting the special nature of the canonical phase measurement \cite{special}. 
For example, it is (up to a unitary equivalence) the only phase shift covariant POVM which generates the number shifts \cite{JP}. 
Hence the number operator and the canonical phase form a canonical pair, similarly to the position and momentum.

It is interesting to note that if one projects the canonical phase POVM $\Ecan$ to a finite dimensional space then the obtained
POVM is
\begin{equation}
\Ecan_l(X)=\sum_{n,m=0}^{l-1}\frac{1}{2\pi}\int_Xe^{i(n-m)\theta}\mathrm{d}\theta \ \kb{n}{m} \, .
\end{equation}
Since $\Ecan_l$ is a continuous outcome POVM on a finite dimensional Hilbert space, it cannot be pure \cite{ChDASc}. 
This is also easy to see directly: for example $\Ecan_l(X)=\frac{1}{2}\Po_l^+(X)+\frac{1}{2}\Po_l^-(X)$, where
\begin{equation*}
\Po_l^\pm(X)=\sum_{n,m=0}^{l-1}\frac{1}{2\pi}\int_X[1\pm\cos(l\theta)]e^{i(n-m)\theta}\mathrm{d}\theta \ \kb{n}{m} \, .
\end{equation*}
It is easy to see that $\Po_l^+$ and $\Po_l^-$ are POVMs on an $l$-dimensional Hilbert space.
Obviously, they are positive since $1\pm\cos(l\theta)\ge 0$. The normalization $\Po_l^\pm([0,2\pi))=\sum_{n=0}^{l-1}\kb n n$
follows from 
\begin{eqnarray*}
&&
\int_0^{2\pi}\cos(l\theta)e^{i(n-m)\theta}\mathrm{d}\theta \\
&&
=
\frac12\int_0^{2\pi}e^{i(n-m+l)\theta}\mathrm{d}\theta+
\frac12\int_0^{2\pi}e^{i(n-m-l)\theta}\mathrm{d}\theta=0
\end{eqnarray*}
since $n-m\pm l\ne 0$ for all $n,m=0,1,...,l-1$.
The infinite dimensional limits $\lim_{l\to\infty}\Po_l^\pm$ do not exist due to the fact that
$\lim_{l\to\infty}\cos(l\theta)$ exists only when $\theta=0$.

On the other hand, if one also discretizes $\Ecan_l$ so that one gets the Pegg-Barnett phase measurements \cite{LaPe}, then the resulting POVM is sharp and hence pure.

Finally, we note that the method used in the proof of Theorem \ref{lause} is not specific only to the canonical phase measurement. 
We expect that with some modifications it is actually applicable to a wide class of continuous outcome POVMs.
The problems related to continuous outcome POVMs in infinite dimension are, typically, rather technical.
Our proof is, on the other hand, quite elementary, therefore showing that with appropriate methods problems can become tractable.    
The structure of pure measurements in infinite dimension is still an open problem, but we hope that the method presented here is a step towards a full characterization.

In conclusion, we have proved that the canonical phase measurement is pure. This result has two important implications.
First of all, this fact demonstrates that in infinite dimension not all pure measurements are either sharp or discrete (nor closely related to them).
Secondly, it shows that the canonical phase measurement is indeed canonical and cannot be interpreted as a noisy measurement, even if it is not a PVM.
This contradicts the folk ``wisdom'' that the difference between PVMs and general POVMs can be explained in terms of (classical) noise.

\par {\em Acknowledgments.---} T. H. acknowledges the financial support by QUANTOP.


\begin{thebibliography}{99}
\bibitem{sigma} Technically, $\Omega$ is a locally compact space and we
  restrict to subsets $X\subseteq\Omega$ which belong to the Borel
  $\sigma$-algebra.
\bibitem{shift} 
Phase shifters are represented by the one-parameter unitary group $e^{i\theta N}$ generated by the number operator $N$.
It was shown in \cite{LaPe} that if a POVM is phase shift covariant in coherent states, then it is actually covariant in all states. 
Therefore, a POVM $\Po$ is phase shift covariant if $e^{i\theta N}\Po(X)e^{-i\theta N} = \Po(X+\theta)$ for all parameters $\theta$ and sets $X$.
\bibitem{decomposition} Infinite and continuous convex decompositions are needed as it is usually preferable to find a convex decomposition in terms of pure constituents. 
Here, however, we are only studying the problem whether a given POVM is pure or not.
\bibitem{special} A comperhensive list of special properties of the canonical phase is gathered in p.\ 51 of \cite{dis}.
\end{thebibliography}
\end{document}